\documentclass[
   ,final            
  ]
  {aipproc}

\usepackage{graphicx}
\layoutstyle{8x11double}
\newcommand{\mysubsection}[1]{  {\bf #1}  }
\begin{document}

\title{Tuning of Kilopixel Transition Edge Sensor Bolometer Arrays with a Digital Frequency
Multiplexed Readout System }

\classification{95.55.Rg, 95.75.Tv, 95.85.Bh, }
\keywords      {bolometers, squids, tuning, parallelization}

\author{Kevin MacDermid}{
  address={McGill University, 3600 rue University, Montreal, QC H3A 2T8}
}
\author{Peter Hyland}{
  address={McGill University, 3600 rue University, Montreal, QC H3A 2T8}
}
\author{Francois Aubin}{
  address={McGill University, 3600 rue University, Montreal, QC H3A 2T8}
}
\author{Eric Bissonnette}{
  address={McGill University, 3600 rue University, Montreal, QC H3A 2T8}
}
\author{Matt Dobbs}{
  address={McGill University, 3600 rue University, Montreal, QC H3A 2T8}
}
\author{Johannes Hubmayr}{
  address={University of Minnesota, 116 Church Street S. E., Minneapolis, MN, 55455}
}
\author{Graeme Smecher}{
  address={McGill University, 3600 rue University, Montreal, QC H3A 2T8}
}
\author{Shahjahan Warraich}{
  address={McGill University, 3600 rue University, Montreal, QC H3A 2T8}
}

\begin{abstract}
A digital frequency multiplexing (DfMUX) system has been developed and used to tune large arrays of transition edge sensor (TES) bolometers read out with SQUID arrays for mm-wavelength cosmology telescopes. The DfMUX system multiplexes the input bias voltages and output currents for several bolometers on a single set of cryogenic wires. Multiplexing reduces the heat load on the camera's sub-Kelvin cryogenic detector stage. In this paper we describe the algorithms and software used to set up and optimize the operation of the bolometric camera. The algorithms are implemented on soft processors embedded within FPGA devices operating on each backend readout board. The result is a fully parallelized implementation for which the setup time is independent of the array size.
\end{abstract}

\maketitle


\section{Introduction}
\label{sec-introduction}
A new generation of mm-wavelength experiments use (or will use) arrays of hundreds or thousands of transition edge sensor (TES) bolometers operating at sub-Kelvin temperatures to image the cosmic microwave background (CMB) to unprecedented precision in the search for new galaxy clusters or the polarization imprint left by inflation. One key element in the cryogenic design for these experiments is the reduction of readout wiring, which is accomplished for APEX-SZ~\cite{2008JLTP..151..697M}, the South Pole Telescope (SPT)~\cite{2004astro.ph.11122R}, EBEX \cite{2008SPIE.7020E..68G} and Polarbear~\cite{2008AIPC.1040...66L} by multiplexing the bolometer sky-signals in the frequency domain.

In this paper we describe tuning algorithms and software developed for the new digital backend electronics~\cite{2008ITNS...55...21D} in use for EBEX and POLARBEAR. This system is the evolution of the analog backend~\cite{2005ApPhL..86k2511L} deployed on APEX-SZ and the SPT. The new system provides an order of magnitude reduction in power consumption and size, improvements in low-frequency noise performance, and a faster, more robust setup. As most of the system complexity is encapsulated in reprogrammable firmware, it is easily adapted to the requirements of specific experiments.

\begin{figure}[htbp]
\includegraphics[width=7cm]{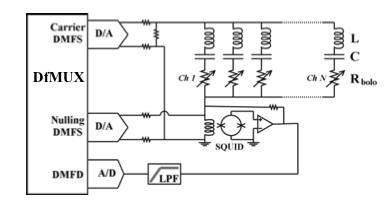} 
\caption{Schematic diagram showing $N$ bolometers multiplexed through a single set of wires and SQUID in the Digital frequency domain multiplexer system.}
\label{fig-schematic}
\end{figure}

Figure \ref{fig-schematic} shows a circuit with $N$-bolometers multiplexed on a single set of wires and read out through a SQUID operating in shunt feedback~\cite{2008ITNS...55...21D, 2005ApPhL..86k2511L}. The box on the left labelled `DfMUX' (for digital frequency multiplexer') represents firmware implemented on a Field Programmable Gate Array (FPGA) that performs the signal processing for the system digitally. A digital multifrequency synthesizer (DMFS) produces a sine-wave bias carrier for each channel and sums them to form a ``comb'' in frequency space. Each bolometer in the multiplexer module is in series with an inductor-capacitor (LC) filter that selects a single bias that is uniquely positioned in frequency-space.

Intensity variations from the sky-signal change the bolometer resistance, amplitude modulating the bolometer current, and appearing as side-bands adjacent to the carriers. These currents are summed at the input of the SQUID, which operates as a transimpedance amplifier. To avoid flux-burdening the SQUID with the large bias carriers, an inverted copy of the carrier ``comb,'' called the nuller, is injected at the SQUID input from a second DMFS. The sky-signal modulated comb is digitized after the SQUID circuit and each bolometer signal is separately demodulated in the digital multifrequency demodulator (DMFD), implemented in firmware.

Each detector channel has its own demodulator, which provides only the in-phase ``I'' component. Including a second demodulator for each channel doubles the usage of FPGA fabric, increasing the power consumption, and does not provide any useful additional information. To align the demodulator phase with the carrier, accounting for phase shifts in the cryostat, each module has an extra $N^{th}+1$ demodulator that can be used to momentarily measure the out-of-phase ``Q'' component of each detector. In addition to aligning the demodulator with the carrier, this ``helper'' demodulator is employed to provide extra information in the algorithms described below.

Frequency domain multiplexing permits the reduction of cryogenic wires from two per bolometer to two per $N$ bolometers, where $N$ is the multiplexing factor. The system is currently implemented with $N=8$ and $N=16$.  Extension to much higher multiplexing factors is feasible but requires a substantial improvement in SQUID bandwidth, using methods such as those demonstrated in Ref.~\cite{2009arXiv0901.1919L}.

\section{Tuning}
\label{sec-tuning}
Tuning the DfMUX system involves, (1) bringing the TES detectors above their superconducting transition and heat-cycling the SQUID to remove trapped flux, (2) tuning the SQUID to its optimum operating point and closing its feedback loop, (3) placing a large sinusoidal bias voltage across the bolometers to keep them normal while the temperature of detector's thermal bath is lowered well below the transition, (4) nulling the bias carriers at the input of the SQUID with out-of-phase sinusoids, (5) lowering the TES into its transition by reducing the bias voltage, and (6) re-nulling to account for the changed bolometer impedance. The algorithms that accomplish these tasks are described below.



\mysubsection{SQUID Tuning}
\label{sec-squid}
Tuning the SQUIDs involves finding the operating points that allow for optimum performance. Small fabrication irregularities, device impurities, and temperature gradients across the cryogenic system necessitate dedicated bias parameter tuning for each device. 

Before tuning the SQUIDs, they are heated above their superconducting transition by injecting a current at heater resistors that are located next to each SQUID in the cryostat. This releases any magnetic flux that may be trapped by the superconducting loops.
After heating, it takes about 10 minutes for the SQUIDs to cool to a stable operating temperature. The bolometer stage is kept at 600~mK, just above the TES transition, by separate heater resistors near the focal plane, or in some cases through optical loading.

\begin{figure}[htbp]
\includegraphics[width=6.0cm]{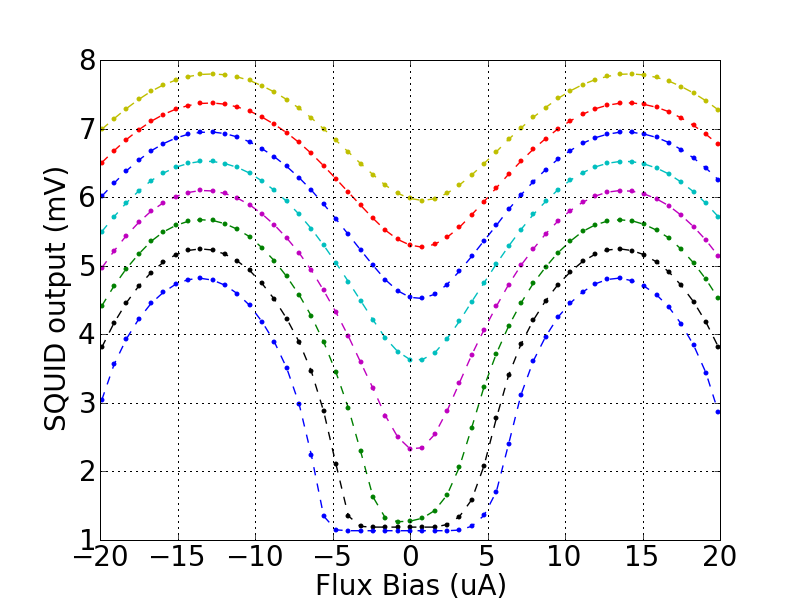} 
\caption{SQUID response curves at increasing SQUID biases from 80 to 115~$\mu A$ . Notice that the peak-to-peak amplitude decreases with increasing SQUID bias (once the curves are above the lower rail) but that the curves become more sinusoidal.}
\label{fig-vphi}
\end{figure}

The voltage response of the open-loop SQUIDs \cite{NIST-arrays} as a function of input current is shown in Figure~\ref{fig-vphi}. Each curve is taken with a distinct current through the SQUID junctions, called the SQUID bias.
The first step of SQUID tuning is determining the optimal SQUID bias, which is a compromise between high peak-to-peak response (indicating a high transimpedance that will provide more loop gain and lower noise in the flux locked loop circuit) and a symmetric sinusoidal response function which results in stable, low-distortion SQUID operation for large input signals once feedback is applied. The algorithm measures the peak-to-peak response as a function of bias current by first measuring one entire response curve at a SQUID bias suspected to be near the maximum. The flux biases of the maximum and minimum SQUID response are determined. The SQUID bias range is then scanned with three measurements taken near the flux bias minimum and three near the maximum. A parabolic fit is used to interpolate, determining the minima, maxima, and peak-to-peak amplitude of each SQUID bias point. The SQUID bias is tuned to the point above the largest response with an amplitude of 90\% of the maximum peak-to-peak. This has been empirically determined to be a good compromise between transimpedance and large signal linearity.

The second step is to adjust the flux bias such that the SQUID is operating on the inverting edge (to effect negative feedback) at the point where the dynamic range is maximized corresponding to half-way between the two extrema. This point is found by tracing out the SQUID response curve at the chosen SQUID bias point.

With both the SQUID bias and flux bias tuned, the SQUID is locked in shunt-feedback by closing a switch connecting a feedback resistor from the output of a low-noise room temperature op-amp following the SQUID to the SQUID input inductor. The feedback serves to linearize the SQUID and extend its dynamic range, resulting in greatly increased SQUID stability.

\mysubsection{Bolometer Overbiasing}
\label{sec-overbias}
Before the bolometer stage can be lowered to its $\sim 250$~mK operating temperature, a large bias is applied at each bolometer frequency to keep the bolometers in their normal state. The required frequencies for these biases are determined in advance by scanning the comb with a carrier test signal. 

Overbiasing is essential, as the voltage bias provided by the backend electronics is not sufficiently large to drive a latched superconducting bolometer normal.

\mysubsection{Carrier Nulling}
\label{sec-nulling}
The bias carrier amplitudes are several orders of magnitude larger than the sky signals and would create a substantial flux burden on the SQUID, pushing it out of its linear regime, if they were not removed at the SQUID input. This removal is accomplished by injecting a ``nulling comb'' at the SQUID input, constructed from sinusoids that are inverted versions of the individual carriers. Because the carrier's current at the SQUID input depends on the bolometer operating point and cryostat wiring, it is necessary to adjust the phase and amplitude of each channel's nulling sinusoid to achieve good results.

\begin{figure}[htbp]
\includegraphics[width=4.5cm]{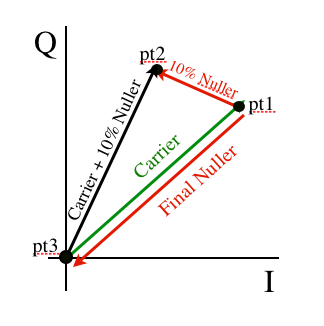} 
\caption{A diagram of the nulling procedure. The first point, pt1, is measured with only the carrier present. The second, pt2, includes a calibration nuller. The difference is used to calculate the phase and amplitude adjustments required to compensate for the carrier.}
\label{fig-nulling}
\end{figure}

Each carrier within a module is nulled one at a time after the bias is applied to the TES detector. In addition to the demodulator associated with each detector channel, the ``helper'' demodulator mentioned in the Introduction section is used to obtain a complex $I$, $Q$ measurement of both amplitude and phase. Note that the phase of the demodulator is arbitrary with respect to the carrier. The algorithm proceeds as follows. First the carrier alone is measured, corresponding to pt1 in Figure~\ref{fig-nulling}. Second, a nuller-calibration point is obtained by injecting a small nulling sinusoid (10\% of the carrier setting) at arbitrary phase and measuring the amplitude and phase of the combined carrier and nuller signal, corresponding to pt2 in Figure~\ref{fig-nulling}. This provides a relative measurement of the nuller's amplitude and phase with respect to the carrier, allowing the calculation of the corrections required to properly null the carrier and return to the origin, or pt3 on Figure~\ref{fig-nulling}. This method has been very successful, the bias carrier is typically nulled to below 1\% of its initial amplitude, allowing recovery of the sky-signal sidebands. 

With the carrier phase information, the channel's demodulator phase is adjusted to be coincident with the carrier, allowing the detector signal to be recovered with a single $I$-demodulation.  The carrier and demodulator use the same system clock, so their relative phase will not wander.

\begin{figure}[htbp]
\includegraphics[width=6.0cm]{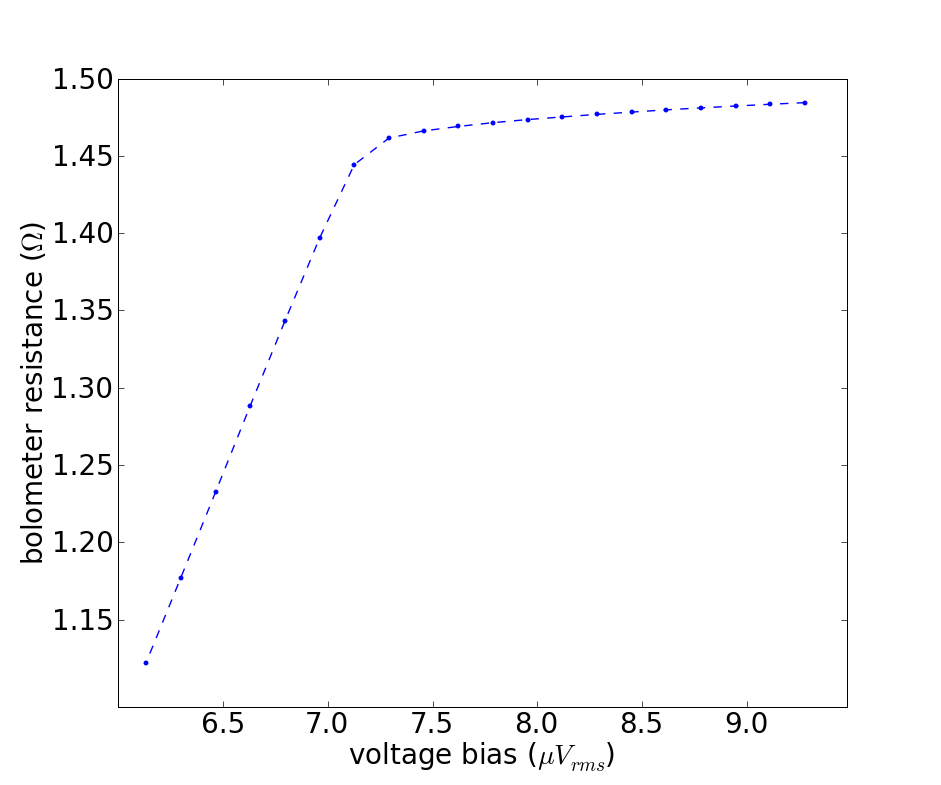} 
\caption{The TES resistance as a function of voltage bias is shown for a typical bolometer during the tuning process.}
\label{fig-bolor}
\end{figure}

\mysubsection{Bolometer Tuning}
\label{sec-bolotuning}
With the bolometers overbiased and the bias carriers nulled, the temperature of the bolometer stage is lowered to its operating point, which takes about an hour. During this period the bolometers are held normal by the electrical power provided. The bolometers are then lowered into their transitions one at a time as follows. First, the relevant nuller is temporarily removed. Then, the carrier's bias voltage is ramped down, while the current is measured at the demodulator. The ramping is stopped when the TES's effective resistance reaches a value defined by the bolometer's stability parameters,  typically 50-80\% of its normal resistance.  Figure \ref{fig-bolor} shows typical TES resistance versus voltage bias curve derived from this procedure. With the TES in its transition, a variant of the nulling procedure is repeated, and the algorithm moves to the next bolometer in the module.

\section{Software Architecture}
In addition to implementing the DSP firmware that synthesizes the waveforms and demodulates the sky signals, the FPGA also houses a soft-core processor (Xilinx Microblaze, www.xilinx.com) running the $\mu$Clinux (Petalogix PetaLinux, www.petalogix.com) operating system, a port of Linux for embedded processors. This allows each DfMUX backend electronics board to manage and execute the software algorithms for the multiplexer modules it handles. All interfaces, including data transport to the experiment's central control computers, is by ethernet.
The embedded processor also provides a simple web-based interface for easy configuration and lab testing. System configuration changes such as running a tuning algorithm or setting carrier frequencies can be performed by sending TCP/IP commands or interactively through the webserver.

The advantage of using web-standard interfaces for the DfMUX board is twofold. First, it enables cross-platform compatibility and, second, it permits higher level scripts to interact with the board in the same fashion regardless of whether they are run on a separate machine or the FPGA itself. Thus, the development path for new algorithms is that they are first written, debugged, and tested on a remote computer executing low-level commands through the TCP/IP interface and later ported to the embedded processor's RAM disk where they can be run in parallel. A custom library, written in the Python scripting language, provides the interface to the system for both low-level commands and high level algorithms \cite{kev-thesis}. This library is the product of several years of development and can reliably tune bolometer arrays in minutes.

A key element in the scalability of the DfMUX system is the ability to parallelize the tuning of the system. This is facilitated by a web-server present on the embedded processor called the DfMUX Algorithm Manager (dalgman).  To request a given algorithm the user sends a string with the desired algorithm name and arguments over ethernet, and is given a key in return. The algorithm then runs on the embedded processor, and stores its data on the DfMUX RAM disk for later retrieval by a user supplying the proper key. Since each backend electronics board has its own embedded processor, the tuning is truly parallel and modular.

The typical time required for setup and tuning of a DfMUX module of bolometers and SQUID is tabulated in the table below. Since the optimum SQUID bias is independent of observing conditions, the SQUID tuning time reported assumes the SQUID bias is known, so only the flux bias must be determined. A full SQUID tuning, including determining the SQUID bias, takes about 140 seconds.

\begin{table}[htbp]  
   \begin{tabular}{ c  c } 
      Script  & Time (s) \\
      \hline
      SQUID tuning & 25 \\
      Overbiasing + Nulling & 142 \\
      Bolometer Tuning + Renull & 76 \\
      \hline
      Total & 243 \\
   \end{tabular}
   \label{benchmarks}
\end{table}

Since the tuning of multiple modules occurs in parallel, this will also be the time required for an array of any size. In principle, once the optimum parameter is found for each device and observing condition (such as bolometer optical loading and orientation of the SQUIDs in the earth's magnetic field), this value could be stored and programmed directly. However, since the tuning time is much shorter than the 2-3 hours required by the cryogenics, and can happen in parallel, further optimization of the tuning time is unwarranted.

\section{Conclusion and Acknowledgments}
The algorithms used to setup and tune large arrays of TES bolometers for observations of the Cosmic Microwave Background radiation with mm-wavelength experiments using a digital frequency domain SQUID multiplexer system are described. The system provides substantial performance and practical improvements over previous generation analog systems.


This work is supported by the Natural Sciences and Engineering Research Council of Canada, 
Canadian Institute for Advanced Research, Canada Research Chairs program, a NASA GSRP grant,
a Minnesota Graduate School Dissertation Fellowship, and the Canadian Foundation for Innovation. 
We thank Xilinx Canada for their contribution of the FPGA devices.



\bibliographystyle{aipproc}   

\bibliography{MyCMB-References}

\IfFileExists{\jobname.bbl}{}
 {\typeout{}
  \typeout{******************************************}
  \typeout{** Please run "bibtex \jobname" to optain}
  \typeout{** the bibliography and then re-run LaTeX}
  \typeout{** twice to fix the references!}
  \typeout{******************************************}
  \typeout{}
 }

\end{document}